\documentclass[intlimits,twoside,a4paper]{article}

\usepackage[cp1251]{inputenc}

\usepackage{multirow,hhline}
\usepackage{mathrsfs}
\usepackage[eqsecnum]{cmpj3}

\issue{2025}{28}{4}{43702}
\doinumber{10.5488/CMP.28.43702}
\title[Absorption and scattering of light by metal-dielectric nanoeggs]%
{Absorption and scattering of light by metal-dielectric nanoeggs}
\author[R. Yu.~Korolkov, R. O.~Malysh, A. V.~Korotun, R. A.~Kulykovskyi]{R. Yu.~Korolkov\orcid{0000-0001-5501-4600}\refaddr{label1},
        R. O.~Malysh\orcid{0009-0001-6086-7381}\refaddr{label1},
        A.~V.~Korotun\orcid{0000-0003-4165-2788}\refaddr{label1,label2}\thanks{Corresponding author: \email{andko@zp.edu.ua}.},
        R. A.~Kulykovskyi\orcid{0000-0001-8781-2113}\refaddr{label1}}
\addresses{
\addr{label1} Zaporizhzhia Polytechnic National University, 69011 Zaporizhzhia, Ukraine
\addr{label2} G. V. Kurdyumov Institute for Metal Physics of the NAS of Ukraine, 03142 Kyiv, Ukraine
}


\date{Received July 20, 2025, in final form August 23, 2025}

\begin{document}

\maketitle

\begin{abstract}
The optical and plasmonic properties of metal-dielectric nanoeggs were investigated in this study. Frequency dependencies of polarizability, absorption and scattering cross-sections, and radiation efficiency were determined. Expressions describing the size-dependent behavior of surface plasmon resonance frequencies were derived. The causes of blue and red shifts in the maxima of polarizability, absorption, and scattering cross-sections as well as variations in their number and amplitude were identified. Recommendations were proposed regarding the use of materials with maximum radiation efficiency in different spectral ranges.
%
%
\printkeywords
%
\end{abstract}

\section{Introduction}

The capability of metal-dielectric structures to sustain coherent electron oscillations, known as surface plasmons, is currently being intensively studied \cite{B01,B02,B03,B04}. Recent advancements in plasmonics have expanded both the number and scope of their practical applications. In particular, the interaction of light with metal nanoparticles of various shapes is being actively investigated \cite{B05,B06}. It has been shown that nanoscale plasmonic structures considerably enhance local electromagnetic fields in specific regions near their surface at certain light frequencies, which are determined by the geometry of the nanostructure. The size-dependent properties of nanoparticles \cite{B07,B08,B09} have potential applications in nanophotonics, biophotonics, and biomedicine \cite{B10,B11}, as well as in scanning near-field optical microscopy \cite{B12}, sensing, and spectroscopic measurements \cite{B13}.

Significant interest in plasmonic phenomena is driven by the discovery and development of surface-enhanced Raman scattering (SERS) techniques \cite{B14}. For instance, gold nanospheres are utilized as enhancing surfaces because they can provide intensity enhancements of up to $10^{14}$ times \cite{B08,B15}. These giant enhancements are attributed to the highly localized electromagnetic fields associated with surface plasmon resonances (SPR). The distribution of scattered light intensity in the near field around a nanostructure can provide valuable insights into its optical properties, which are crucial for various applications.

The frequency of SPR, determined by the dielectric function of the metal and the permittivity of the surrounding environment, strongly depends on the size and shape of the nanostructure \cite{B16,B17}. Wet chemical synthesis methods have enabled the fabrication of plasmonic nanoparticles with a variety of shapes, such as spheres \cite{B18}, triangular plates \cite{B19,B20}, rods~\cite{B21} and cubes~\cite{B22} with controlled sizes and narrow size distributions. Additionally, nanoparticles of ``metal--dielectric'' and ``metal--metal'' core--shell types, as well as mixed alloy nanoparticles of various shapes (e.g., nanoshells~\cite{B23,B24,B25} and nanorice \cite{B26}), have been fabricated.

The capability of producing such a wide variety of nanostructures opens the door to numerous applications, since the spectral position of the SPR depends on both the shape and size of the nanoparticle. For spherical nanoparticles, the resonance of oscillating electrons with the incident optical wave occurs when the real part of the dielectric permittivity of the particle equals the negative of twice the dielectric permittivity of the surrounding medium \cite{B27}. However, for nonspherical particles, electron oscillations are anisotropic and are localized either along the main axes \cite{B28}, or at the edges and corners of the nanoparticle \cite{B29}, This leads to an additional shape-dependent depolarization and the splitting of SPR into two modes --- longitudinal and transverse modes in nanorods, nanodiscs, and biconical structures~\cite{B30,B31,B32} or symmetric and antisymmetric modes in nanoshells \cite{B33,B34}.

 Of particular interest are the optical properties of a hybrid (in particular, metal--dielectric) core--shell (A@B) nanoparticles, where a core made of material A is coated with a shell of material B. This interest arises from the fact that such structures provide additional opportunities for controlling surface plasmon resonance, such as varying the core and shell materials and their relative sizes to tailor the optical response for specific applications.

Applications of nanoplasmonics, such as SERS \cite{B35}, photothermal tumor therapy \cite{B36} and thermoplasmonics \cite{B37} require a considerably enhancement of the near field in the vicinity of nanoparticles. Nanoshells with uniform thickness are less attractive in this respect because reducing the thickness of the metal layer, necessary to tune the localized SPR peak position, weakens the near fields. Consequently, for these applications, structures such as rough surfaces \cite{B38} and nanoparticle assemblies \cite{B39} are generally preferrable. The advantage of these structures lies in their capability to concentrate light into small volumes, considerably enhance the local electric field in these regions, and create the so-called ``hot spots''~\cite{B40}. However, these structures have also got a significant drawback: the hot spots can be located in hard-to-access areas, such as interparticle gaps or within surface irregularities.

For this reason, researchers utilize non-concentric spherical nanoparticles (nanoeggs) and misaligned cylinders in which the inner dielectric core is displaced relative to the center or axis of the metallic shell~\cite{B41,B42,B43,B44}. These types of nanoshells offer an additional degree of freedom for tuning the SPR frequency by varying the core offset \cite{B41}. Experimental studies have demonstrated that these nanoshells are promising candidates for applications such as surface-enhanced Raman spectroscopy, photothermal tumor therapy, nanoscale optical sensing, and photothermal agents for a selective destruction of antibiotic-resistant bacteria \cite{B41,B42,B43,B44,B45}.

The plasmonic properties of concentric nanoshells in both near and far fields have been studied using the quasistatic approach, finite element methods (FEM), finite-difference time-domain (FDTD) methods, and discrete dipole approximation (DDA) \cite{B46}. Systematic investigations have explored the effects of size, the dielectric permittivity of the core and nanoshell, and the shell thickness on the optical and plasmonic properties \cite{B34,B46}. Additionally, the optical properties of two-layered spherical nanoparticles, both metal-dielectric and bimetallic, have been studied in detail \cite{B47,B48,B49,B50,B51}. Furthermore, numerical modelling has been conducted for isolated non-concentric dielectric nanoshells with metallic cores, revealing significant differences in their plasmonic properties compared to isolated concentric nanoshells \cite{B41,B42,B43,B44}. However, the optical properties of structures with a dielectric core and a metallic shell of variable thickness (nanoeggs) remain largely unexplored. Thus, addressing this gap is a highly relevant research objective.

\begin{figure}[htb]
	\centerline{\includegraphics[scale=0.55]{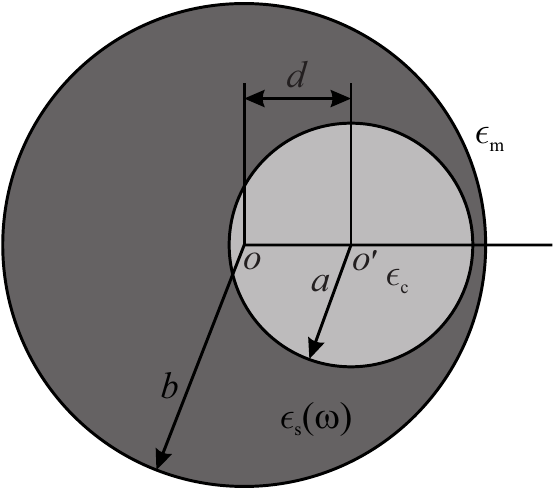}}
	\caption{Geometry of the problem.} \label{fig1}
\end{figure}

\section{Mathematical model}
\subsection{Polarizability, absorption, and scattering cross-sections of a nanoegg}

Let us consider the interaction of light with a nanoegg embedded in a dielectric medium with permittivity ${\epsilon_{\rm{m}}}$. The studied structure consists of two non-concentric nanospherical components with radii $a$ and $b$, and a distance $d$ between their centers. The dielectric permittivities of the core and the shell materials are ${\epsilon_{\rm{c}}}$ and ${\epsilon_{\rm{s}}}$, respectively (figure~\ref{fig1}).

For each sphere, we introduce local spherical coordinate systems: $\left( {r,\,\phi ,\,\theta } \right)$ for the outer shell and $\left( {r',\,\phi ',\,\theta '} \right)$ for the inner core.

Since the wavelength of the incident light is much larger than the size of the system ($\lambda  \gg a,\,b,\,d$), the problem can be addressed within the quasistatic approximation. In this approximation, the electric potential in each region [1 --- core (c), 2 --- shell (s), 3 --- surrounding the dielectric medium (m)] is determined by solving Laplace’s equations:
\begin{align}
\label{eq1}
\Delta {\varphi _i} = 0,\quad (i = {\rm{c}},\,{\rm{s}},\,{\rm{m}}),
\end{align}
with the following boundary conditions:
\begin{align}
{\left. {{\varphi _{\rm{c}}}} \right|_{r' = a}} &= {\left. {{\varphi _{\rm{s}}}} \right|_{r' = a}},
\nonumber\\
{\left. {\epsilon_{\rm{c}}\frac{{\partial {\varphi _{\rm{c}}}}}{{\partial r'}}} \right|_{r' = a}} &= {\left. {{\epsilon_{\rm{s}}}\frac{{\partial {\varphi _{\rm{s}}}}}{{\partial r'}}} \right|_{r' = a}},
\nonumber\\
{\left. {{\varphi _{\rm{s}}}} \right|_{r = b}} &= {\left. {{\varphi _{\rm{m}}}} \right|_{r = b}},
\nonumber\\
{\left. {{\epsilon_{\rm{s}}}\frac{{\partial {\varphi _{\rm{s}}}}}{{\partial r}}} \right|_{r = b}} &= {\left. {{\epsilon_{\rm{m}}}\frac{{\partial {\varphi _{\rm{m}}}}}{{\partial r}}} \right|_{r = b}}.
\label{eq2}
\end{align}
The solutions to Laplace’s equations for all three regions can be expressed as:
\begin{align}
{\varphi _{\rm{c}}}\left( {r',\,\,\theta '} \right)& = \sum\limits_{n = 1}^\infty  {{A_n}{{\left( {\frac{{r'}}{a}} \right)}^n}{P_n}\left( {\cos \theta '} \right)} ,
\nonumber\\
{\varphi _{\rm{s}}}\left( {r,\,\,\theta } \right) &= \sum\limits_{n = 1}^\infty  {\left[ {{B_n}{{\left( {\frac{r}{b}} \right)}^n} + {C_n}{{\left( {\frac{b}{r}} \right)}^{n + 1}}} \right]{P_n}\left( {\cos \theta } \right)} ,
\nonumber\\
{\varphi _{\rm{m}}}\left( {r,\,\,\theta } \right) &= \sum\limits_{n = 1}^\infty  {\left[ {{D_n}{{\left( {\frac{b}{r}} \right)}^n} + {E_n}{{\left( {\frac{r}{b}} \right)}^{n + 1}}} \right]{P_n}\left( {\cos \theta } \right)},
\label{eq3}
\end{align}
where
\begin{equation}
\label{eq4}
{E_n} = \left\{ \begin{array}{l}
 -\mathscr{E}_{0}b,\,\quad n = 1,\\
\,\,\,\,\,0,\qquad n \geqslant 2,
\end{array} \right.
\end{equation}
and $\mathscr{E}_0$ is the electric field of the incident light wave; ${P_n}\left( x \right)$ are Legendre polynomials.

After straightforward transformations, the following relation for the dimensionless polarizability of the nanoegg is obtained (see appendix \ref{appendix_A}):
\begin{align}
{{\tilde \alpha }_@} =\, &1 - 3{\epsilon_{\rm{m}}}\Bigg\{ { - 2{\beta _{\rm{c}}}\left( {{\epsilon_{\rm{s}}} - {\epsilon_{\rm{m}}}} \right)\left( {{\epsilon_{\rm{s}}} - {\epsilon_{\rm{c}}}} \right) + \left( {{\epsilon_{\rm{s}}} + 2{\epsilon_{\rm{m}}}} \right)}  
\nonumber\\
&\times {\left. {\left[ {{\epsilon_{\rm{c}}} + 2{\epsilon_{\rm{s}}} + \frac{{6{\beta _{\rm{}}}\delta _{\rm{c}}^{{2 \mathord{\left/
 {\vphantom {2 3}} \right.
 \kern-\nulldelimiterspace} 3}}\left( {{\epsilon_{\rm{s}}} - {\epsilon_{\rm{c}}}} \right)\left( {{\epsilon_{\rm{s}}} - {\epsilon_{\rm{m}}}} \right)\left( {3{\epsilon_{\rm{s}}} + 2{\epsilon_{\rm{c}}}} \right)}}{{ - \frac{1}{2}\left( {2{\epsilon_{\rm{s}}} + 3{\epsilon_{\rm{m}}}} \right)\left( {3{_{\rm{s}}} + 2{\epsilon_{\rm{c}}}} \right) + 3\beta _{\rm{c}}^{{5 \mathord{\left/
 {\vphantom {5 3}} \right.
 \kern-\nulldelimiterspace} 3}}\left( {{\epsilon_{\rm{s}}} - {\epsilon_{\rm{c}}}} \right)\left( {{\epsilon_{\rm{s}}} - {\epsilon_{\rm{m}}}} \right)}}} \right]} \right\}^{ - 1}} 
\nonumber\\
&\times \left[ {{\beta _{\rm{c}}}\left( {{\epsilon_{\rm{s}}} - {\epsilon_{\rm{c}}}} \right) + 2{\epsilon_{\rm{s}}} + {\epsilon_{\rm{c}}} + \frac{{6{\beta _{\rm{c}}}\delta _{\rm{c}}^{{2 \mathord{\left/
 {\vphantom {2 3}} \right.
 \kern-\nulldelimiterspace} 3}}\left( {{\epsilon_{\rm{s}}} - {\epsilon_{\rm{c}}}} \right)\left( {{\epsilon_{\rm{s}}} - {\epsilon_{\rm{m}}}} \right)\left( {3{\epsilon_{\rm{s}}} + 2{\epsilon_{\rm{c}}}} \right)}}{{ - \frac{1}{2}\left( {2{\epsilon_{\rm{s}}} + 3{\epsilon_{\rm{m}}}} \right)\left( {3{\epsilon_{\rm{s}}} + 2{\epsilon_{\rm{c}}}} \right) + 3\beta _{\rm{c}}^{{5 \mathord{\left/
 {\vphantom {5 3}} \right.
 \kern-\nulldelimiterspace} 3}}\left( {{\epsilon_{\rm{s}}} - {\epsilon_{\rm{c}}}} \right)\left( {{\epsilon_{\rm{s}}} - {\epsilon_{\rm{m}}}} \right)}}} \right],
\label{eq5}
\end{align}
where the ``@'' index refers to a core--shell structure of the type ``A@B'', in which the dielectric core A is coated with a metallic shell B;
\begin{equation}
\label{eq6}
{\beta _{\rm{c}}} = \frac{{{a^3}}}{{{b^3}}},
\quad
{\delta _{\rm{c}}} = \frac{{{d^3}}}{{{b^3}}}
\end{equation}
represent the volumetric fraction of the dielectric in the nanostructure and the relative displacement of the centers of the dielectric core and the entire nanoparticle, respectively.

In the limiting case of ${\delta _{\rm{c}}} \to 0$, the expression for the dimensionless polarizability of a nanoshell with constant thickness is obtained (see, for example, \cite{B51})
$$
{\tilde \alpha _@} = \frac{{\left( {{\epsilon_{\rm{s}}} - {\epsilon_{\rm{m}}}} \right)\left( {2{\epsilon_{\rm{s}}} + {\epsilon_{\rm{c}}}} \right) - {\beta _{\rm{c}}}\left( {{\epsilon_{\rm{s}}} - {\epsilon_{\rm{c}}}} \right)\left( {2{\epsilon_{\rm{s}}} + {\epsilon_{\rm{m}}}} \right)}}{{\left( {{\epsilon_{\rm{s}}} + 2{\epsilon_{\rm{m}}}} \right)\left( {2{\epsilon_{\rm{s}}} + {\epsilon_{\rm{c}}}} \right) - 2{\beta _{\rm{c}}}\left( {{\epsilon_{\rm{s}}} - {\epsilon_{\rm{c}}}} \right)\left( {{\epsilon_{\rm{s}}} - {\epsilon_{\rm{m}}}} \right)}}.
$$
Since the core material is dielectric, then ${\epsilon_{\rm{c}}} = {\mathop{\rm const}\nolimits} $, and the dielectric function of the metal is described by the Drude model
\begin{equation}
\label{eq7}
{\epsilon_{\rm{s}}} = {\epsilon^\infty } - \frac{{\omega _p^2}}{{\omega \left( {\omega  + {\mathop{\rm i}\nolimits} {\gamma _{{\rm{eff}}}}} \right)}},
\end{equation}
where ${\omega _p}$ is the plasma frequency (${\omega _p} = \sqrt {{{n_e}{e^2}}/{{\epsilon_0}{m^*}}}$; $\epsilon_0$ is the dielectric constant; ${n_e}$ and ${m^*}$ are the concentration and effective mass of electrons, respectively; $n_e^{ - 1} = {\frac43 \piup r_s^3}$ and $r_s$ is the average distance between the conduction electrons); ${\epsilon^\infty }$ is the lattice contribution to the dielectric function, and ${\gamma _{{\rm{eff}}}}$ is the effective relaxation rate, the size dependence of which will be determined below.

The absorption and scattering cross-sections are determined by the following relations:
\begin{align}
C_@^{{\rm{abs}}} = \frac{\omega }{c}V\sqrt {{\epsilon_{\rm{m}}}} \Im {{\tilde \alpha }_@},
\nonumber\\
C_@^{{\rm{sca}}} = \frac{{{\omega ^4}}}{{{c^4}}}{V^2}\epsilon_{\rm{m}}^2{\left| {{{\tilde \alpha }_@}} \right|^2},
\label{eq8}
\end{align}
where the volume of the nanoegg is $V = \frac43 \piup {b^3} $, and the radiation efficiency is given by
\begin{equation}
\label{eq9}
\xi _@^{{\rm{rad}}} = \frac{1}{{1 + \big({{C_@^{{\rm{abs}}}}}/{{C_@^{{\rm{sca}}}}}\big)}}.
\end{equation}

\subsection{Effective relaxation rate}

The effective relaxation rate is assumed to include additive contributions from three energy loss mechanisms: bulk relaxation, surface relaxation, and radiative damping:
\begin{equation}
\label{eq10}
{\gamma _{{\rm{eff}}}} = {\gamma _{{\rm{bulk}}}} + {\gamma _{\rm{s}}} + {\gamma _{{\rm{rad}}}},
\end{equation}
where the expressions for the components of the relaxation rate are as follows (see appendix \ref{appendix_B}):
\begin{description}
\item[--] Bulk relaxation rate
\begin{equation}
\label{eq11}
{\gamma _{{\rm{bulk}}}} = \frac{1}{{\left[ {2 - {{\left( {1 - \frac{{\beta _{\rm{c}}^{{2 \mathord{\left/
 {\vphantom {2 3}} \right.
 \kern-\nulldelimiterspace} 3}}}}{{1 - \delta _{\rm{c}}^{{2 \mathord{\left/
 {\vphantom {2 3}} \right.
 \kern-\nulldelimiterspace} 3}}}}} \right)}^{{1 \mathord{\left/
 {\vphantom {1 2}} \right.
 \kern-\nulldelimiterspace} 2}}}} \right]{\tau _{{\rm{bulk}}}}}}.
\end{equation}
\item[--] Surface relaxation rate
\begin{align}
{\gamma _{\rm{s}}} =\,& \frac{{{v_{\rm{F}}}}}{b}\left( {1 + \beta _{\rm{c}}^{{2 \mathord{\left/
 {\vphantom {2 3}} \right.
 \kern-\nulldelimiterspace} 3}}} \right)\left\{ {1 - \frac{{\beta _{\rm{c}}^{{2 \mathord{\left/
 {\vphantom {2 3}} \right.
 \kern-\nulldelimiterspace} 3}}}}{{{{\left( {1 - \delta _{\rm{c}}^{{1 \mathord{\left/
 {\vphantom {1 3}} \right.
 \kern-\nulldelimiterspace} 3}}} \right)}^2}}} + \frac{1}{2}\left( {1 - \delta _{\rm{c}}^{{1 \mathord{\left/
 {\vphantom {1 3}} \right.
 \kern-\nulldelimiterspace} 3}}} \right)\left( {1 + \beta _{\rm{c}}^{{2 \mathord{\left/
 {\vphantom {2 3}} \right.
 \kern-\nulldelimiterspace} 3}}} \right)\left( {1 - \frac{{\beta _{\rm{c}}^{{1 \mathord{\left/
 {\vphantom {1 3}} \right.
 \kern-\nulldelimiterspace} 3}}}}{{1 - \delta _{\rm{c}}^{{1 \mathord{\left/
 {\vphantom {1 3}} \right.
 \kern-\nulldelimiterspace} 3}}}}} \right)} \right. 
\nonumber\\
&- {\left. {\frac{1}{4}\left( {1 - \delta _{\rm{c}}^{{1 \mathord{\left/
 {\vphantom {1 3}} \right.
 \kern-\nulldelimiterspace} 3}}} \right)\left[ {1 - \frac{{\beta _{\rm{c}}^{{2 \mathord{\left/
 {\vphantom {2 3}} \right.
 \kern-\nulldelimiterspace} 3}}}}{{{{\left( {1 - \delta _{\rm{c}}^{{1 \mathord{\left/
 {\vphantom {1 3}} \right.
 \kern-\nulldelimiterspace} 3}}} \right)}^2}}} + \beta _{\rm{c}}^{{1 \mathord{\left/
 {\vphantom {1 3}} \right.
 \kern-\nulldelimiterspace} 3}}\frac{{{{\left( {1 - \delta _{\rm{c}}^{{1 \mathord{\left/
 {\vphantom {1 3}} \right.
 \kern-\nulldelimiterspace} 3}}} \right)}^2} - \beta _{\rm{c}}^{{2 \mathord{\left/
 {\vphantom {2 3}} \right.
 \kern-\nulldelimiterspace} 3}}}}{{1 - \delta _{\rm{c}}^{{1 \mathord{\left/
 {\vphantom {1 3}} \right.
 \kern-\nulldelimiterspace} 3}}}}} \right]\ln \frac{{1 - \beta _{\rm{c}}^{{1 \mathord{\left/
 {\vphantom {1 3}} \right.
 \kern-\nulldelimiterspace} 3}} - \delta _{\rm{c}}^{{1 \mathord{\left/
 {\vphantom {1 3}} \right.
 \kern-\nulldelimiterspace} 3}}}}{{1 + \beta _{\rm{c}}^{{1 \mathord{\left/
 {\vphantom {1 3}} \right.
 \kern-\nulldelimiterspace} 3}} - \delta _{\rm{c}}^{{1 \mathord{\left/
 {\vphantom {1 3}} \right.
 \kern-\nulldelimiterspace} 3}}}}} \right\}^{ - 1}}.
\label{eq12}
\end{align}
\item[--] Radiative damping rate
\begin{equation}
\label{eq13}
{\gamma _{{\rm{rad}}}} = \frac{2}{9}\frac{V}{{\epsilon_{\rm{m}}^{{1 \mathord{\left/
 {\vphantom {1 2}} \right.
 \kern-\nulldelimiterspace} 2}}}}\left( {1 - {\beta _{\rm{c}}}} \right){\left( {\frac{{{\omega _p}}}{c}} \right)^3} \cdot \left\{ \begin{array}{l}
\,\,\,{\gamma _{\rm{s}}},\,\,\,\,\,\,\,\ell _{{\rm{bulk}}}^s > 2{t_{{\rm{eff}}}},\\
{\gamma _{{\rm{bulk}}}},\,\,\,\,\ell _{{\rm{bulk}}}^s \leqslant 2{t_{{\rm{eff}}}}.
\end{array} \right.
\end{equation}
\end{description}
In the limiting case, from expressions (\ref{eqB16}) and (\ref{eqB18}) we obtain the effective electron mean free path and the shell thickness for a system of two concentric spheres:
\begin{equation}
\label{eq14}
\mathop {\lim }\limits_{{\delta _{\rm{c}}} \to 0} {L_{{\rm{eff}}}} = \frac{{b\Big( {1 - \beta _{\rm{c}}^{{1}/{3}}} \Big)}}{{1 + \beta _{\rm{c}}^{{2}/{3}}}}\left[ {1 + \beta _{\rm{c}}^{{1}/{3}} - \frac{1}{2}\Big( {1 + \beta _{\rm{c}}^{{2}/{3}}} \Big) - \frac{1}{4}{{\Big( {1 + \beta _{\rm{c}}^{{1}/{3}}} \Big)}^2}\ln \frac{{1 - \beta _{\rm{c}}^{{1}/{3}}}}{{1 + \beta _{\rm{c}}^{{1}/{3}}}}} \right],
\end{equation}
\begin{equation}
\label{eq15}
\mathop {\lim }\limits_{{\delta _{\rm{c}}} \to 0} {t_{{\rm{eff}}}} = b\left( {1 - \beta _{\rm{c}}^{{1 \mathord{\left/
 {\vphantom {1 3}} \right.
 \kern-\nulldelimiterspace} 3}}} \right),
\end{equation}
which agrees with the results from \cite{B52}.

To obtain numerical results, we use expressions (\ref{eq5}), (\ref{eq8}) and (\ref{eq9}) incorporating formulas~(\ref{eq6}), (\ref{eq7}), (\ref{eq10})--(\ref{eq13}) and (\ref{eqB18}).

\section{Calculation results and their discussion}

Calculations of the frequency dependencies of polarizability, absorption and scattering cross-sections, as well as radiation efficiency, were performed for nanoeggs with various geometric parameters, shells made of different metals, and cores composed of different dielectrics. The parameters of the metallic shells and dielectric cores are presented in tables \ref{tab1} and \ref{tab2}, respectively. The surrounding medium was assumed to be teflon (${\epsilon_{\rm{m}}} = 2.3$).

\begin{table}[htb]
	\caption{Parameters of metals ($z$ is the valency of metals, $a_0$ is the Bohr radius) \cite{B04}.}
	\label{tab1}
	\vspace{1ex}
	\centering
	\begin{tabular}{|c|c|c|c|c|c|}
		\hline
		\multirow{2}{*}{Metal} & \multicolumn{5}{c|}{Parameter} \\
		\hhline{~-----} & $z$ & ${r_s}/{a_0}$ & ${m^*}/{m_e}$ & ${\epsilon^\infty }$ & ${\gamma _{{\rm{bulk}}}},\,\,{10^{13}}\,\,{{\rm{s}}^{ - 1}}$ \\ \hline
		Pd & 2 & 4.00 & 0.37 & 2.52 & 13.9 \\ \hline
		Pt & 2 & 3.27 & 0.54 & 4.42 & 10.52 \\ \hline
		Ag & 1 & 3.02 & 0.96 & 3.70 & 2.50 \\ \hline
		\multirow{2}{*}{Au} & 1 & 3.01 & \multirow{2}{*}{0.99} & \multirow{2}{*}{9.84} & \multirow{2}{*}{3.45}\\
		\hhline{~--~~~}\multirow{2}{*}{} & 3 & 2.09 & \multirow{2}{*}{} & \multirow{2}{*}{} & \multirow{2}{*}{}\\ \hline
		\multirow{2}{*}{Cu} & 1 & 2.67 & \multirow{2}{*}{1.49} & \multirow{2}{*}{12.03} & \multirow{2}{*}{3.45}\\
		\hhline{~--~~~}\multirow{2}{*}{} & 2 & 2.11 & \multirow{2}{*}{} & \multirow{2}{*}{} & \multirow{2}{*}{}\\ \hline
		\multirow{3}{*}{Al} & \multirow{3}{*}{3} & \multirow{3}{*}{2.07} & 1.06 & \multirow{3}{*}{0.7} & \multirow{3}{*}{1.25}\\
		\hhline{~~~-~~}\multirow{3}{*}{} & \multirow{3}{*}{} & \multirow{3}{*}{} & 1.48 & \multirow{3}{*}{} & \multirow{3}{*}{}\\
		\hhline{~~~-~~}\multirow{3}{*}{} & \multirow{3}{*}{} & \multirow{3}{*}{} & 1.60 & \multirow{3}{*}{} & \multirow{3}{*}{}\\ \hline
	\end{tabular}
\end{table}

\begin{table}[htb]
	\caption{Dielectric permittivities of core materials \cite{B53}.}
	\label{tab2}
	\vspace{1ex}
	\centering
	\begin{tabular}{|c|c|c|c|c|c|}
		\hline
		Core & ${\rm{Si}}{{\rm{O}}_2}$ & ${\rm{A}}{{\rm{l}}_2}{{\rm{O}}_3}$ & ${\rm{ZnO}}$ & ${\rm{T}}{{\rm{a}}_2}{{\rm{O}}_5}$ & ${\rm{N}}{{\rm{b}}_2}{{\rm{O}}_5}$ \\ \hline
		${\epsilon_{\rm{c}}}$ & 2.10 & 3.10 & 4.00 & 4.67 & 6.15 \\ \hline
	\end{tabular}
\end{table}

\begin{figure}[!h]
	\centerline{\includegraphics[scale=0.48]{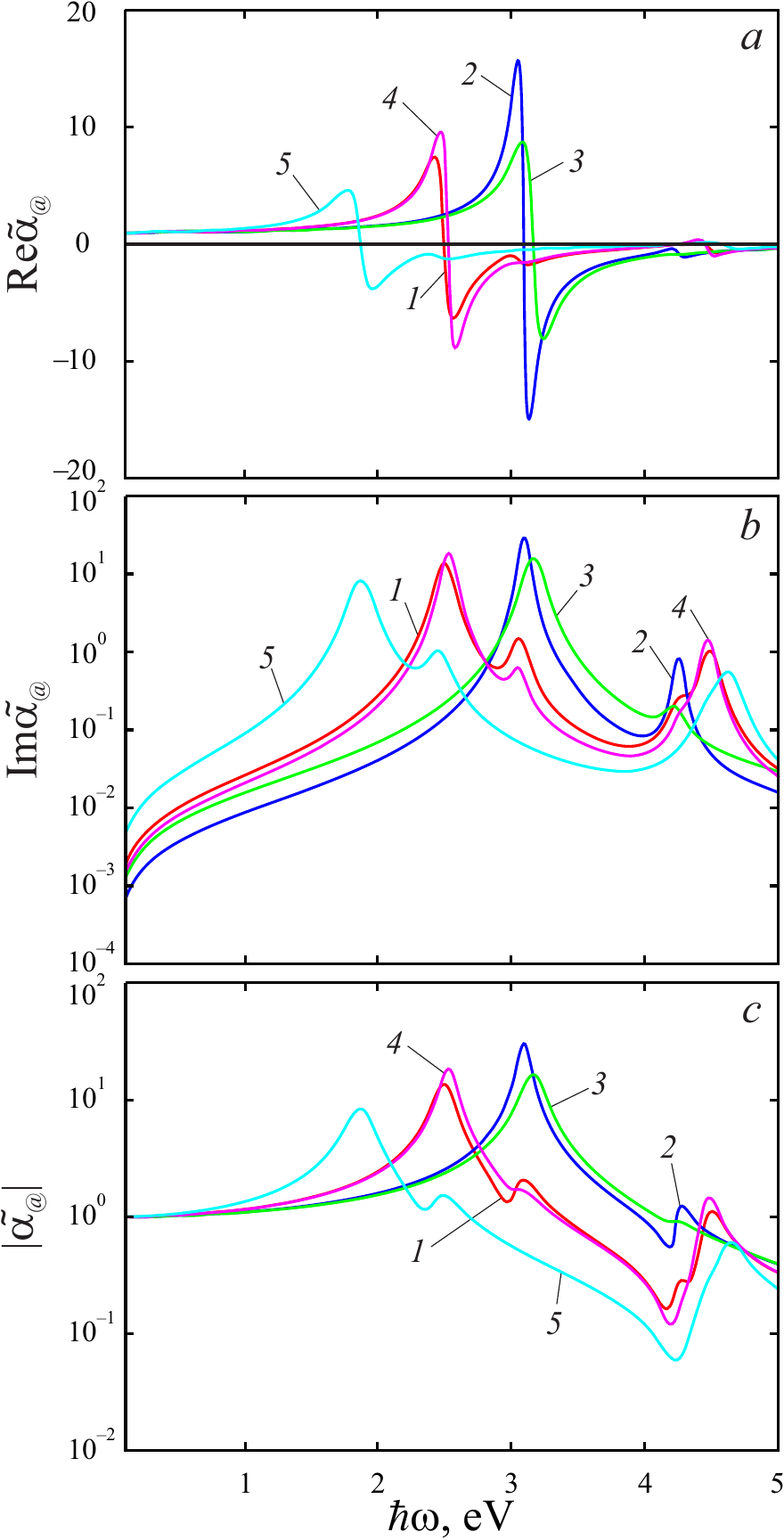}}
	\caption{(Colour online) Frequency dependencies of the real (a) and imaginary (b) parts, as well as the modulus (c) of the dimensionless polarizability for nanoeggs   with different inner and outer radii (${\rm{Si}}{{\rm{O}}_2}@{\rm{Ag}}$) $d = 2\,\,{\rm{nm}}$: 1 ---  $a = 10\,\,{\rm{nm}}$, $b = 15\,\,{\rm{nm}}$; 2 --- $a = 10\,\,{\rm{nm}}$, $b = 30\,\,{\rm{nm}}$; 3 --- $a = 10\,\,{\rm{nm}}$, $b = 50\,\,{\rm{nm}}$; 4 --- $a = 20\,\,{\rm{nm}}$, $b = 30\,\,{\rm{nm}}$; 5 --- $a = 25\,\,{\rm{nm}}$, $b = 30\,\,{\rm{nm}}$.} \label{fig2}
\end{figure}

Figure \ref{fig2} presents the calculated frequency dependencies of the real and imaginary parts, as well as the modulus of the nanoegg polarizability ${\rm{Si}}{{\rm{O}}_2}@{\rm{Ag}}$ for various core and overall particle radii while maintaining a fixed distance between their centers. Similar to metallic, metal-dielectric, and bimetallic nanoparticles of different shapes \cite{B47,B48,B49,B50,B51,B52}, the real part of the polarizability is a sign-changing function of frequency, whereas the imaginary part is positive throughout the investigated frequency range. When the overall particle radius increases (while keeping the core radius constant) $\max \left\{ {\Im \tilde{\alpha} } \right\}$ exhibits a ``blue'' shift (curves in the sequence $1 \to 2 \to 3$). Conversely, an increase in the core radius (with a fixed outer radius) results in a shift toward lower frequencies (curves in the sequence $2 \to 4 \to 5$). Thus, a reduction in the size of the dielectric core and an increase in the outer radius of the nanoparticle corresponding to a higher metal content leads to a ``blue'' shift of the first maxima of ${\Im \tilde{\alpha} }$, which are associated with lower-frequency surface plasmon resonances. It is worth noting that subsequent maxima of $\max \left\{ {\Im \tilde{\alpha} } \right\}$ have low amplitudes and are primarily located in the near-ultraviolet frequency range. The splitting of surface plasmon resonances occurs due to two main factors:		
\begin{enumerate}
	\item The presence of two ``metal-dielectric'' interfaces (core--shell and shell--surrounding medium).
	\item The non-concentric nature of the spheres.
\end{enumerate}	

Since the other resonances (besides the first one) have low amplitudes and occur at closely spaced frequencies, they appear in figure~\ref{fig2} as small, closely positioned peaks in the blue and ultraviolet spectral ranges. Given that $\left| \Re {\tilde \alpha } \right| \sim \Im {\tilde \alpha }$, contribute approximately equally to $\left| {\tilde \alpha } \right|$, the corresponding curves in figure~\ref{fig2}c exhibit the characteristics of both $\Re {\tilde \alpha }$ and $\Im {\tilde \alpha }$.

\begin{figure}[!h]
	\centerline{\includegraphics[scale=0.48]{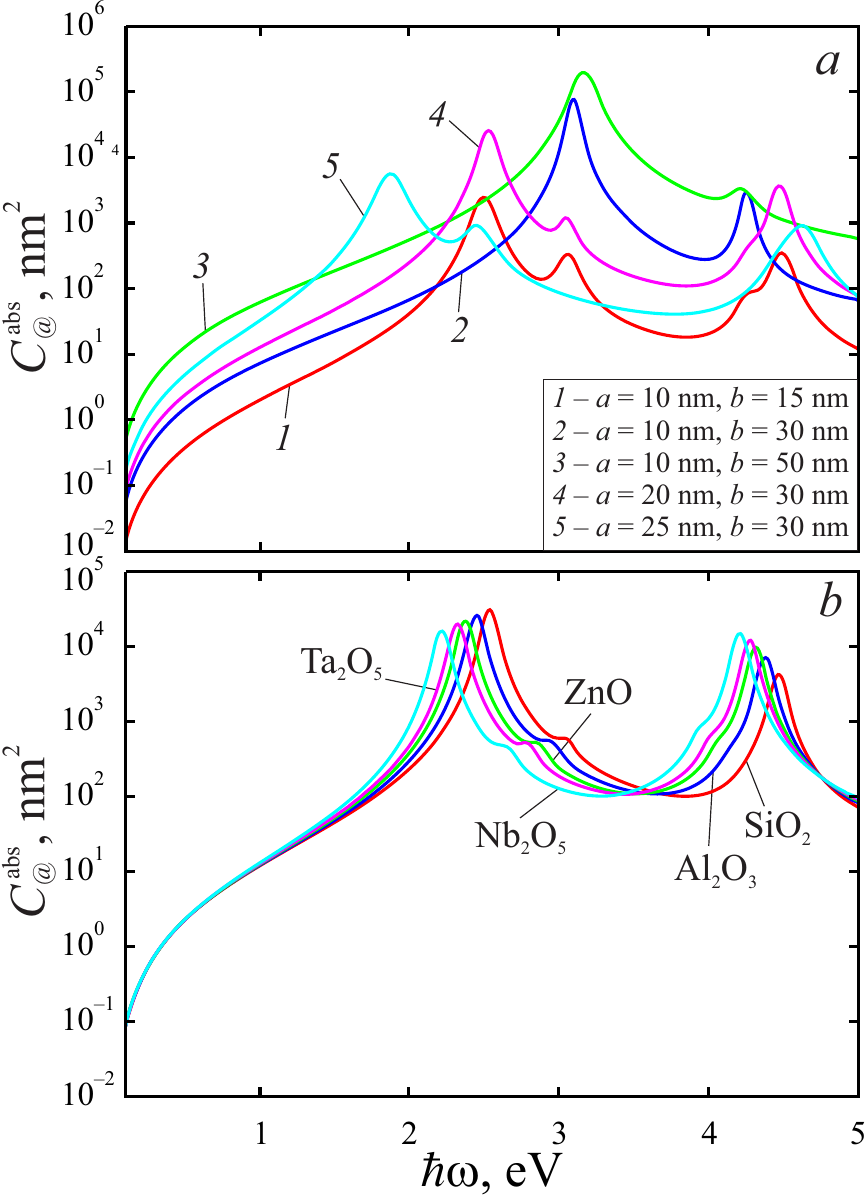}}
	\caption{(Colour online) Frequency dependencies of the absorption cross-sections for nanoeggs ${\rm{Si}}{{\rm{O}}_2}@{\rm{Ag}}$: Nanoeggs with different inner and outer radii (a) and silver nanoshells with different dielectric core materials (b) at parameters: $a = 20\,\,{\rm{nm}}$, $b = 30\,\,{\rm{nm}}$, $d = 2\,\,{\rm{nm}}$.} \label{fig3}
\end{figure}

Now, we derive the equation for determining the frequencies of the SPR. This equation follows from the condition that the denominator of expression (\ref{eq5}) equals zero in the non-dissipative approximation (when ${\gamma _{{\rm{eff}}}} \to 0$) and has the following form:
\begin{align}
&\left\{ {\left( {\epsilon_{\rm{s}}^{sp} + 2{\epsilon_{\rm{m}}}} \right)\left( {2_{\rm{s}}^{sp} + {\epsilon_{\rm{c}}}} \right) - 2{\beta _{\rm{}}}\left( {\epsilon_{\rm{s}}^{sp} - {\epsilon_{\rm{m}}}} \right)\left( {\epsilon_{\rm{s}}^{sp} - {\epsilon_{\rm{c}}}} \right)} \right\} 
\nonumber\\
&\times \left\{ { - \frac{1}{2}\left( {2\epsilon_{\rm{s}}^{sp} + 3{\epsilon_{\rm{m}}}} \right)\left( {3\epsilon_{\rm{s}}^{sp} + 2{\epsilon_{\rm{c}}}} \right) + 3\beta _{\rm{}}^{{5 \mathord{\left/
 {\vphantom {5 3}} \right.
 \kern-\nulldelimiterspace} 3}}\left( {\epsilon_{\rm{s}}^{sp} - {\epsilon_{\rm{}}}} \right)\left( {\epsilon_{\rm{s}}^{sp} - {\epsilon_{\rm{m}}}} \right)} \right\} 
\nonumber\\
&+ 6{\beta _{\rm{}}}\delta _{\rm{}}^{{2 \mathord{\left/
 {\vphantom {2 3}} \right.
 \kern-\nulldelimiterspace} 3}}\left( {\epsilon_{\rm{s}}^{sp} - {\epsilon_{\rm{}}}} \right)\left( {\epsilon_{\rm{s}}^{sp} - {\epsilon_{\rm{m}}}} \right)\left( {3\epsilon_{\rm{s}}^{sp} + 2{\epsilon_{\rm{}}}} \right) = 0,
\label{eq16}
\end{align}
where
\begin{equation}
\label{eq17}
\epsilon_{\rm{s}}^{sp} = {\epsilon^\infty } - \frac{{\omega _p^2}}{{\omega _{sp}^2}}.
\end{equation}

From equations (\ref{eq16}) and (\ref{eq17}), it follows that there are four SPR frequencies, determined by the formula:
\begin{equation}
\label{eq18}
{\omega _{sp}} = \frac{{{\omega _p}}}{{\sqrt {{\epsilon^\infty } - \epsilon_{\rm{s}}^{\left( i \right)}} }},
\end{equation}
where $\epsilon_{\rm{s}}^{\left( i \right)}$ are the solutions of a quartic equation derived from expression (\ref{eq16}):
\begin{equation}
\label{eq19}
{a_4}\epsilon_{\rm{s}}^{sp\,\,4} + {a_3}\epsilon_{\rm{s}}^{sp\,\,3} + {a_2}\epsilon_{\rm{s}}^{sp\,\,2} + {a_1}\epsilon_{\rm{s}}^{sp} + {a_0} = 0,
\end{equation}
\noindent where the coefficients ${a_i}$ take the following form:
\begin{align}
{a_4} =\, &6\left[ {3{\beta _{\rm{c}}}\delta _{\rm{c}}^{{2 \mathord{\left/
 {\vphantom {2 3}} \right.
 \kern-\nulldelimiterspace} 3}} + \left( {1 - {\beta _{\rm{c}}}} \right)\left( {\beta _{\rm{c}}^{{5 \mathord{\left/
 {\vphantom {5 3}} \right.
 \kern-\nulldelimiterspace} 3}} - 1} \right)} \right],
\nonumber\\
{a_3} =\, &6{\beta _{\rm{c}}}\delta _{\rm{c}}^{{2 \mathord{\left/
 {\vphantom {2 3}} \right.
 \kern-\nulldelimiterspace} 3}}\left( {3{\epsilon_{\rm{m}}} - {\epsilon_{\rm{}}}} \right) + 3\left( {\beta _{\rm{c}}^{{5 \mathord{\left/
 {\vphantom {5 3}} \right.
 \kern-\nulldelimiterspace} 3}} - 1} \right)\left[ {2\left( {2 + {\beta _{\rm{c}}}} \right){\epsilon_{\rm{m}}} + \left( {1 + 2{\beta _{\rm{c}}}} \right){\epsilon_{\rm{c}}}} \right] 
\nonumber\\
&- 2\left( {1 - {\beta _{\rm{c}}}} \right)\left[ {\left( {2 + 3\beta _{\rm{}}^{{5 \mathord{\left/
 {\vphantom {5 3}} \right.
 \kern-\nulldelimiterspace} 3}}} \right){\epsilon_{\rm{c}}} + 3\left( {\frac{3}{2} + \beta _{\rm{c}}^{{5 \mathord{\left/
 {\vphantom {5 3}} \right.
 \kern-\nulldelimiterspace} 3}}} \right){\epsilon_{\rm{m}}}} \right],
\nonumber\\
{a_2} = \, &- 2\epsilon_{\rm{}}^2 - {\epsilon_{\rm{c}}}{\epsilon_{\rm{m}}} - 6\epsilon_{\rm{m}}^2 + 12\left( {1 - {\beta _{\rm{c}}}} \right)\left( {\beta _{\rm{}}^{{5 \mathord{\left/
 {\vphantom {5 3}} \right.
 \kern-\nulldelimiterspace} 3}} - 1} \right){\epsilon_{\rm{c}}}{\epsilon_{\rm{m}}} 
\nonumber\\
&-\left[ {2\left( {2 + {\beta _{\rm{c}}}} \right){\epsilon_{\rm{m}}} + \left( {1 + 2{\beta _{\rm{c}}}} \right){\epsilon_{\rm{c}}}} \right]\left[ {\left( {2 + \beta _{\rm{}}^{{5 \mathord{\left/
 {\vphantom {5 3}} \right.
 \kern-\nulldelimiterspace} 3}}} \right){\epsilon_{\rm{}}} + 3\left( {\frac{3}{2} + \beta \epsilon_{\rm{c}}^{{5 \mathord{\left/
 {\vphantom {5 3}} \right.
 \kern-\nulldelimiterspace} 3}}} \right){\epsilon_{\rm{m}}}} \right],
\nonumber\\
{a_1} =\, &{\epsilon_{\rm{c}}}{\epsilon_{\rm{m}}}\left\{ {2\left( {{\epsilon_{\rm{m}}} - {\epsilon_{\rm{}}}} \right) - 2\left( {\beta _{\rm{c}}^{{5 \mathord{\left/
 {\vphantom {5 3}} \right.
 \kern-\nulldelimiterspace} 3}} - 1} \right)\left[ {\left( {2 + 3\beta _{\rm{c}}^{{5 \mathord{\left/
 {\vphantom {5 3}} \right.
 \kern-\nulldelimiterspace} 3}}} \right){\epsilon_{\rm{}}} + 3\left( {\frac{3}{2} + \beta _{\rm{c}}^{{5 \mathord{\left/
 {\vphantom {5 3}} \right.
 \kern-\nulldelimiterspace} 3}}} \right){\epsilon_{\rm{m}}}} \right]} \right. 
\nonumber\\
&+ \left. {3\left( {\beta _{\rm{c}}^{{5 \mathord{\left/
 {\vphantom {5 3}} \right.
 \kern-\nulldelimiterspace} 3}} - 1} \right)\left[ {2\left( {2 + {\beta _{\rm{c}}}} \right){\epsilon_{\rm{m}}} + \left( {1 + 2{\beta _{\rm{c}}}} \right){\epsilon_{\rm{c}}}} \right]} \right\},\nonumber\\
{a_0} =\, &6\left[ {4{\beta _{\rm{c}}}\delta _{\rm{c}}^{{2 \mathord{\left/
 {\vphantom {2 3}} \right.
 \kern-\nulldelimiterspace} 3}} + \left( {1 - {\beta _{\rm{c}}}} \right)\left( {\beta _{\rm{c}}^{{5 \mathord{\left/
 {\vphantom {5 3}} \right.
 \kern-\nulldelimiterspace} 3}} - 1} \right)} \right]\epsilon_{\rm{c}}^2\epsilon_{\rm{m}}^2.
\label{eq20}
\end{align}

\begin{figure}[!h]
	\centerline{\includegraphics[scale=0.43]{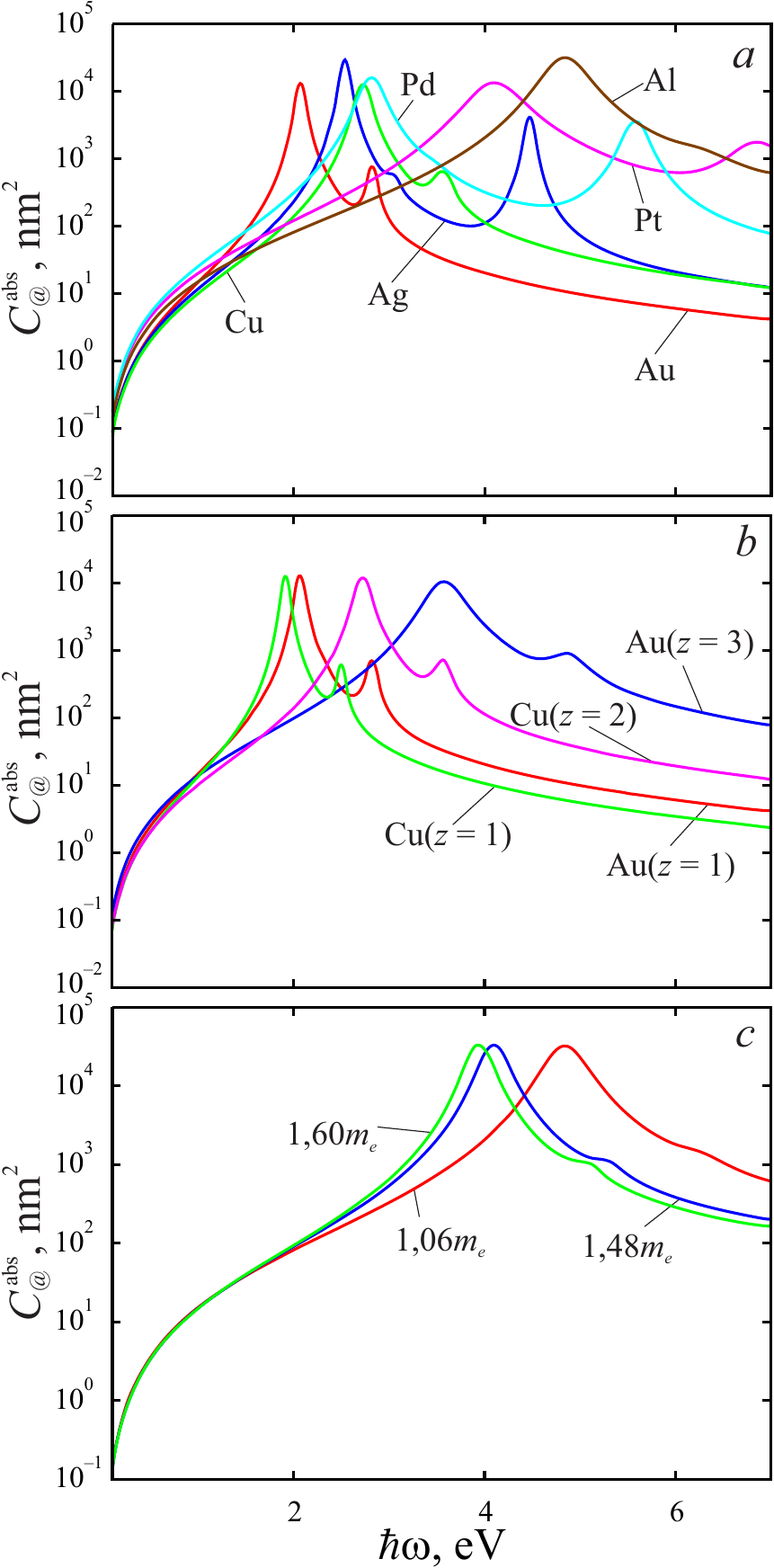}}
	\caption{(Colour online) Frequency dependencies of the absorption cross-sections for nanoeggs: ${\rm{Si}}{{\rm{O}}_2}@{\rm{Me}}$ (\emph{a}), nanoeggs ${\rm{Si}}{{\rm{O}}_2}@{\rm{Au}}$ and  ${\rm{Si}}{{\rm{O}}_2}@{\rm{Cu}}$ at different valency values (\emph{b}), nanoeggs ${\rm{Si}}{{\rm{O}}_2}@{\rm{Al}}$ at different effective electron masses (\emph{c}): $a = 20\,\,{\rm{nm}}$, $b = 30\,\,{\rm{nm}}$, $d = 2\,\,{\rm{nm}}$.} \label{fig4}
\end{figure}

Figure \ref{fig3} presents the frequency-dependent absorption cross-section curves for nanoeggs ${\rm{Si}}{{\rm{O}}_2}@{\rm{Ag}}$ with different inner and outer radii, as well as for particles with various dielectric cores covered by a silver shell. Notably, the number of absorption cross-section maxima varies for different core and nanoparticle radii and depends on the geometric parameter $\frac{d}{b - a}$ (figure~\ref{fig3}a). A decrease in this parameter leads to a reduction in the number of maxima (due to merging), decreasing from four to two when $\frac{d}{b - a} \to 0$ (i.e., for concentric spheres). Additionally, increasing the dielectric permittivity of the core material in the series ${\rm{Si}}{{\rm{O}}_2} \to {\rm{A}}{{\rm{l}}_2}{{\rm{O}}_3} \to {\rm{ZnO}} \to {\rm{T}}{{\rm{a}}_2}{{\rm{O}}_5} \to {\rm{N}}{{\rm{b}}_2}{{\rm{O}}_5}$ results in a slight ``red'' shift of the absorption cross-section maxima (figure~\ref{fig3}b).

Figure \ref{fig4} presents the frequency dependence curves of the absorption cross-sections for different cases: shells made of various metals (figure~\ref{fig4}a), Au and Cu shells, where the atoms of these metals have different valencies (figure~\ref{fig4}b), and Al shells with different effective electron masses (figure~\ref{fig4}c). Since ${\omega_p}$, and consequently, the surface plasmon resonance frequencies increase in the sequence of metals ${\rm{Au}} \to {\rm{Ag}} \to {\rm{Cu}} \to {\rm{Pd}} \to {\rm{Pt}} \to {\rm{Al}}$, the corresponding absorption cross-section maxima (both the first and subsequent ones) follow the same order (figure~\ref{fig4}a). Furthermore, since ${\omega _p}\sim \sqrt {{n_e}}$, and ${n_e} \sim z$, an increase in valency leads to a shift of the absorption cross-section maxima toward higher frequencies~(figure~\ref{fig4}b). With an increase in the effective electron mass, due to the relationship ${\omega _p} \sim {1}/{{\sqrt {{m^*}}}}$ a ``red'' shift of the absorption cross-section maxima occurs for the nanoeggs ${\rm{Si}}{{\rm{O}}_2}@{\rm{Al}}$~(figure~\ref{fig4}c).

\begin{figure}[!h]
	\centerline{\includegraphics[scale=0.45]{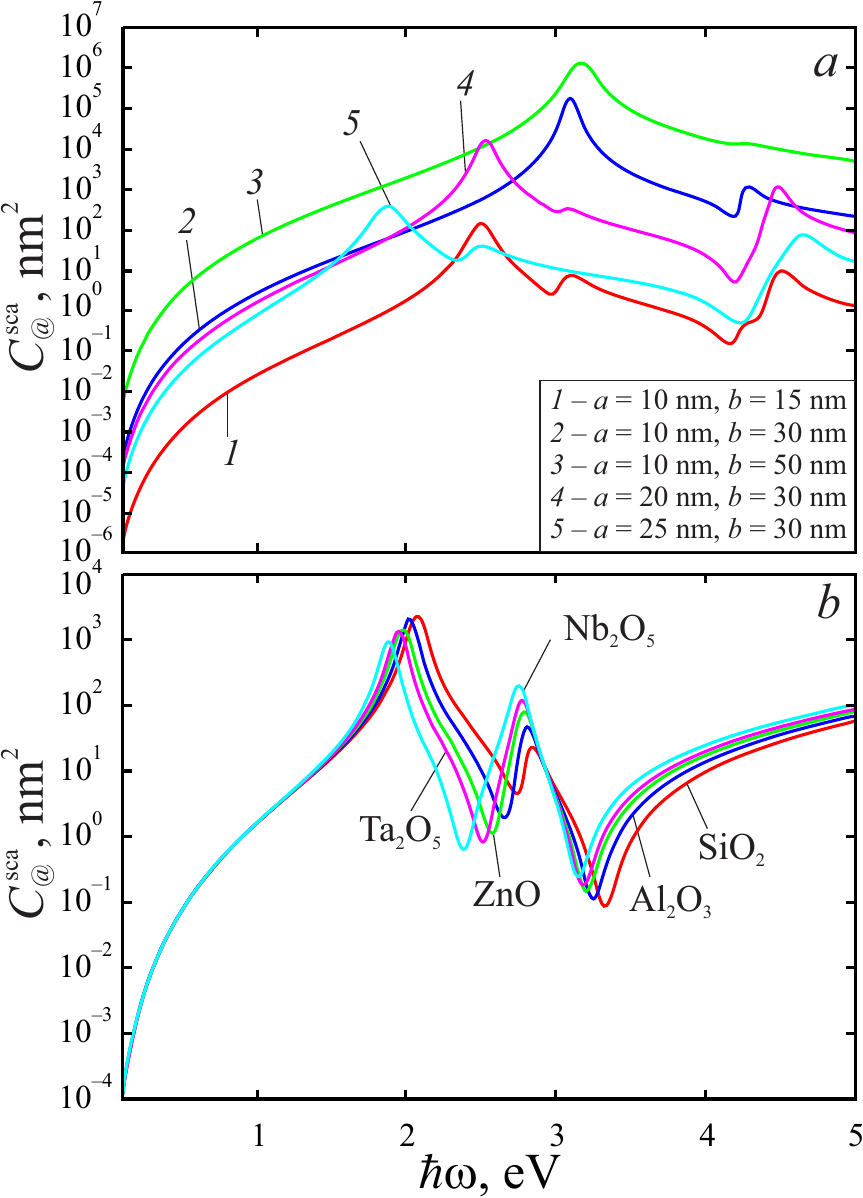}}
	\caption{(Colour online) Frequency dependencies of the scattering cross-sections for nanoeggs ${\rm{Si}}{{\rm{O}}_2}@{\rm{Ag}}$ with different inner and outer radii (\emph{a}) and silver nanoshells with different dielectric cores (\emph{b}) at parameters: $a = 20\,\,{\rm{nm}}$, $b = 30\,\,{\rm{nm}}$, $d = 2\,\,{\rm{nm}}$.} \label{fig5}
\end{figure}

Numerical experiments have shown that the values and the number of real roots of equation~(\ref{eq19}) considerably depend on the coefficients ${a_i}$, and consequently, on the geometric parameters of the nanoeggs, as well as the optical properties of the dielectric cores and metallic shells. For small values of ${\beta _{\rm{c}}}$ (``thin shell'') and ${\delta _{\rm{c}}}$ (low non-concentricity), as well as in the case of gold and copper shells, equation~(\ref{eq19}) has two real roots, meaning that the curves of  $C_@^{{\rm{abs}}}\left( \omega  \right)$ will exhibit two maxima. However, for silver shells, an increase in ${\beta _{\rm{c}}}$ and ${\delta _{\rm{c}}}$ leads to an increase in the number of real roots of equation~(\ref{eq19}), and consequently, the number of absorption cross-section maxima $C_@^{{\rm{abs}}}\left( \omega  \right)$ increases to four.

\begin{figure}[!h]
	\centerline{\includegraphics[scale=0.45]{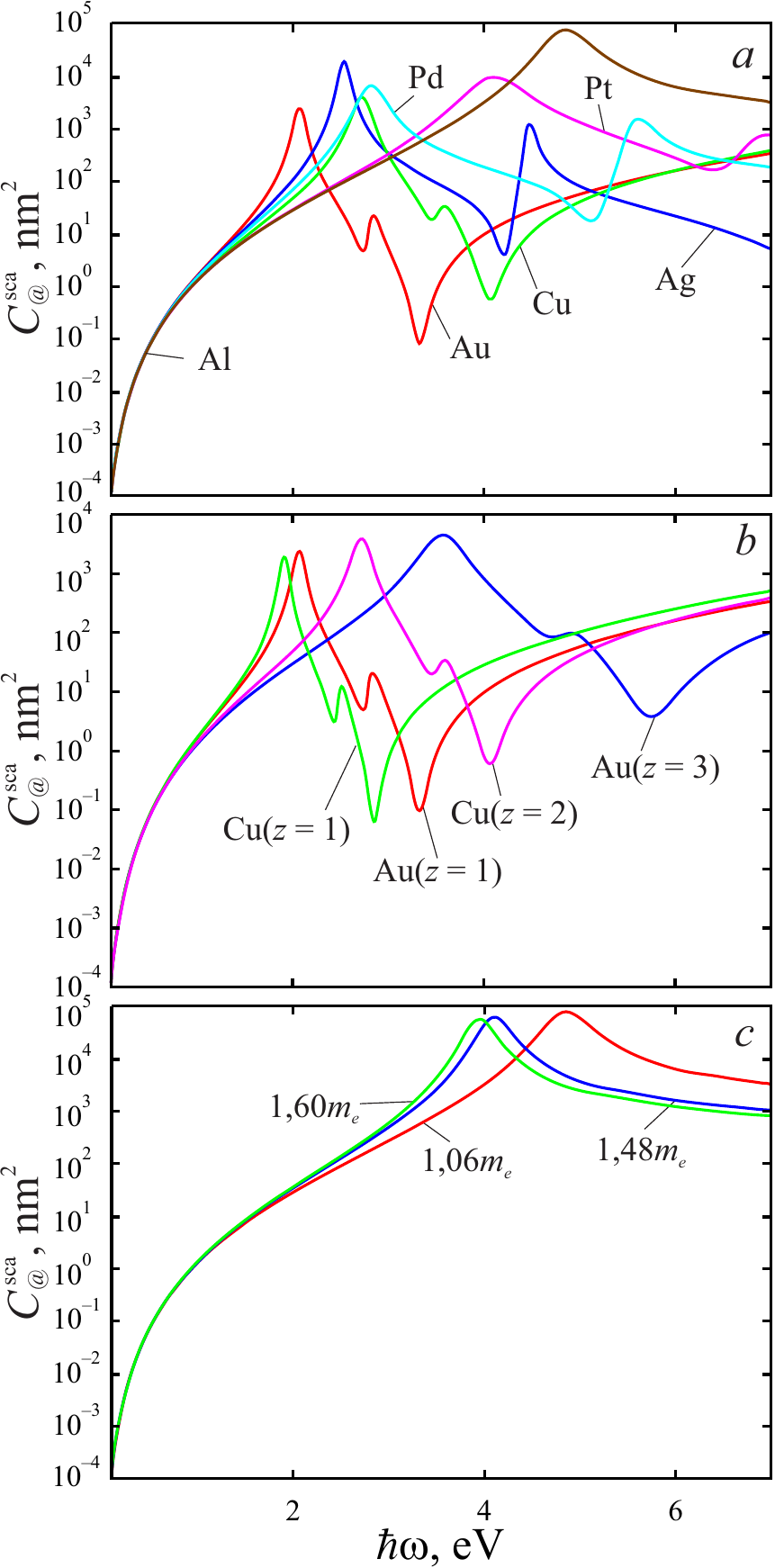}}
	\caption{(Colour online) Frequency dependencies of the scattering cross-sections for nanoeggs: ${\rm{Si}}{{\rm{O}}_2}@{\rm{Me}}$~(\emph{a}), nanoeggs ${\rm{Si}}{{\rm{O}}_2}@{\rm{Au}}$ and ${\rm{Si}}{{\rm{O}}_2}@{\rm{Cu}}$ at different valency values (\emph{b}), nanoeggs ${\rm{Si}}{{\rm{O}}_2}@{\rm{Al}}$ at different effective electron masses (\emph{c}). Parameters: $a = 20\,\,{\rm{nm}}$, $b = 30\,\,{\rm{nm}}$, $d = 2\,\,{\rm{nm}}$.} \label{fig6}
\end{figure}

Analogous to the curves of the frequency dependencies of the absorption cross-sections (figure~\ref{fig3}), the curves for the scattering cross-sections are demonstrated in figure~\ref{fig5}. It should be noted that the number and nature of the shifts in the maxima of the scattering cross-sections fully correspond to the case of absorption cross-sections, both for nanoeggs ${\rm{Si}}{{\rm{O}}_2}@{\rm{Ag}}$ with different core and total particle radii and for particles where cores of different dielectrics are coated with a silver shell. This analogy between the frequency dependence curves of absorption and scattering cross-sections is also observed for cases where the shell is made of different metals (figure~\ref{fig6}a), Au and Cu, when the atoms of these metals have different valencies (figure~\ref{fig6}b), and for Al shells with different effective electron masses (figure~\ref{fig6}c).

\begin{figure}[!h]
\centerline{\includegraphics[scale=0.45]{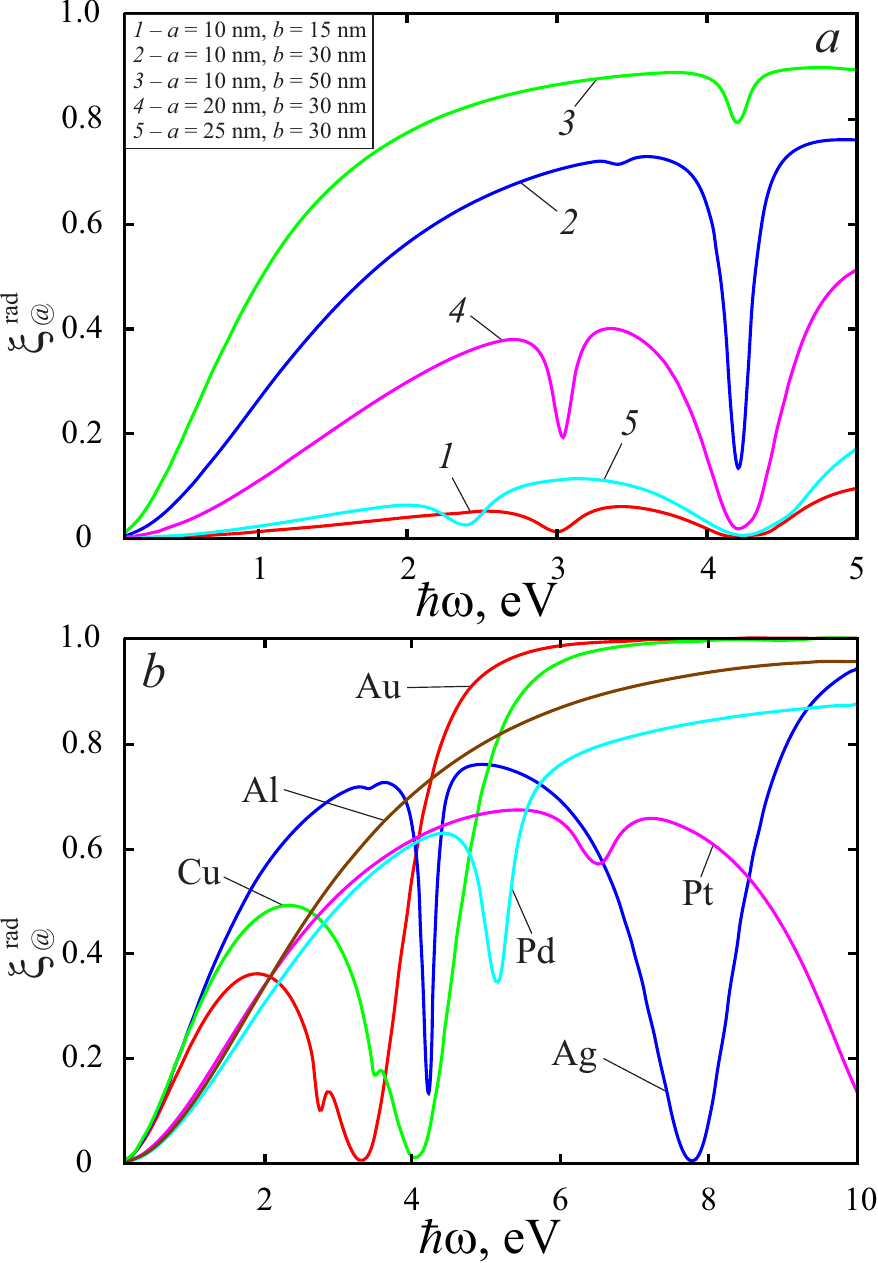}}
\caption{(Colour online) Frequency dependencies of radiation efficiency for nanoeggs ${\rm{Si}}{{\rm{O}}_2}@{\rm{Ag}}$ with different inner and outer radii~(\emph{a}) and for nanoeggs ${\rm{Si}}{{\rm{O}}_2}@{\rm{Me}}$ (\emph{b}). Parameters: $a = 20\,\,{\rm{nm}}$, $b = 30\,\,{\rm{nm}}$, $d = 2\,\,{\rm{nm}}$.} \label{fig7}
\end{figure}

Figure \ref{fig7} presents the frequency dependence curves of radiation efficiency for nanoeggs ${\rm{Si}}{{\rm{O}}_2}@{\rm{Ag}}$ of different sizes and for particles with shells made of different metals. With an increase in the overall radius of ${\rm{Si}}{{\rm{O}}_2}@{\rm{Ag}}$ while keeping the core radius constant (curves in the sequence $1 \to 2 \to 3$) and with a decrease in the dielectric core radius while maintaining a fixed outer radius (curves in the sequence $5 \to 4 \to 2$), an increase in the metal content within the composite particle leads to an increase in radiation efficiency, a reduction in the amplitude of the minima, and a decrease in their number from two to one (figure~\ref{fig7}a). It is interesting to note that while in the visible spectral range, the maximum radiation efficiency for nanoeggs ${\rm{Si}}{{\rm{O}}_2}@{\rm{Ag}}$, in the ultraviolet frequency range, the highest radiation efficiency is observed for nanoeggs ${\rm{Si}}{{\rm{O}}_2}@{\rm{Cu}}$.

In conclusion, we note that in future works it is planned to investigate the optical properties of such systems as a nanoegg on a dielectric substrate and composite media with inclusions of the metal-dielectric nanoegg type.

\section{Conclusions}

Expressions for the frequency dependencies of polarizability, absorption and scattering cross-sections, and radiation efficiency have been obtained for composite nanoparticles of the ``dielectric core -- metallic shell with variable thickness'' type (nanoeggs).

It has been demonstrated that the expressions for polarizability and effective relaxation rate reduce to the corresponding formulas for a metallic shell of constant thickness in the limiting case.

It has been proven that the number and position of the maxima of the imaginary part of polarizability depend on the geometric parameters (core and total nanoparticle radii, distance between the centers of the core and the entire particle).

Specifically, an increase in metal content (by increasing the total particle radius or decreasing the core radius) leads to a ``blue'' shift of the maxima, whereas reducing the distance between the centers of the core and the nanoparticle decreases the number of maxima from four to two (in the case of concentric spheres). All maxima, except for the first one, have low amplitudes and are located in the blue and ultraviolet spectral ranges.

It has been shown that increasing the permittivity of the dielectric core shifts the absorption and scattering cross-section maxima toward lower frequencies.

It has been established that a ``blue'' shift in the absorption and scattering cross-section maxima occurs when the shell metal changes in the sequence from gold to aluminum, when the valency of the metal atoms increases, and when the effective electron mass decreases. This is due to an increase in plasma frequency, and consequently, the frequency of the surface plasmon resonance.

It has been demonstrated that the highest radiation efficiency is achieved for nanoeggs with the highest metal content. Additionally, at identical geometric parameters, the radiation efficiency is influenced by the nature of the shell material: in the visible range, radiation efficiency is maximal for particles with a silver shell, whereas in the near-ultraviolet range, it is the highest for particles with a copper shell.

\appendix
\section{Derivation of the formula for nanoegg polarizability}
\label{appendix_A}

We describe the procedure for obtaining the frequency dependence of polarizability.

Substituting solutions (\ref{eq3}) into the boundary conditions (\ref{eq2}), we obtain:
\begin{align}
&\sum\limits_{n = 1}^\infty  {{A_n}{P_n}\left( {\cos \theta '} \right)}  =\sum\limits_{n = 1}^\infty  {{{\left. {\left[ {{B_n}{{\left( {\frac{r}{b}} \right)}^n} + {C_n}{{\left( {\frac{b}{r}} \right)}^{n + 1}}} \right]{P_n}\left( {\cos \theta } \right)} \right|}_{r' = a}}} ,
\nonumber\\
&{\epsilon_{\rm{}}}\sum\limits_{n = 1}^\infty  {n{A_n}{P_n}\left( {\cos \theta '} \right)}  ={\epsilon_{\rm{s}}}a{\sum\limits_{n = 1}^\infty  {\left. {\left[ {{B_n}{{\left( {\frac{r}{b}} \right)}^n} + {C_n}{{\left( {\frac{b}{r}} \right)}^{n + 1}}} \right]{P_n}\left( {\cos \theta } \right)} \right|} _{r' = a}},
\nonumber\\
&\sum\limits_{n = 1}^\infty  {\left[ {{B_n} + {C_n}} \right]{P_n}\left( {\cos \theta } \right)}  = \sum\limits_{n = 1}^\infty  {\left[ {{D_n} + {E_n}} \right]{P_n}\left( {\cos \theta } \right)} ,
\nonumber\\
&{\epsilon_{\rm{s}}}\left[ {n{B_n} - \left( {n + 1} \right){C_n}} \right]{P_n}\left( {\cos \theta } \right) = {\epsilon_{\rm{m}}}\sum\limits_{n = 1}^\infty  {\left[ {n{E_n} - \left( {n + 1} \right){D_n}} \right]{P_n}\left( {\cos \theta } \right)}.
\label{eqA1}
\end{align}
Using the orthogonality condition:
\begin{equation}
\label{eqA2}
\int {{P_l}\left( {\cos \theta } \right){P_n}\left( {\cos \theta } \right)} \,\rd\left({\cos \theta } \right) = \frac{2}{{2l + 1}}{\delta _{ln}},
\end{equation}
where ${\delta _{ln}}$ is the Kronecker symbol, and the addition theorem for harmonic functions (see, for example,~\cite{B54}) is used
\begin{equation}
\label{eqA3}
r'\strut^{n}{P_n}\left( {\cos \theta '} \right) = \sum\limits_{k = 0}^n {\frac{{n!}}{{k!\left( {n - k} \right)!}}{d^{n - k}}{r^k}{P_k}\left( {\cos \theta } \right)},
\end{equation}
\begin{equation}
\label{eqA4}
r'\strut^{-(n+1)}{P_n}\left( {\cos \theta '} \right) = \sum\limits_{k = 0}^n {{{\left( { - 1} \right)}^{k - n}}\frac{{k!}}{{n!\left( {k - n} \right)!}}{d^{k - n}}\frac{{{P_k}\left( {\cos \theta } \right)}}{{{r^{k + 1}}}}},
\end{equation}
rewrite the system of equations (\ref{eqA1}) in the form:
\begin{align}
&{A_l} = \sum\limits_{n = 1}^\infty  {{K_{ln}}{B_n}}  + \sum\limits_{n = 1}^\infty  {{M_{ln}}{C_n}} ,
\nonumber\\
&{B_l} + {C_l} = {E_l} + {D_l},
\nonumber\\
&l{\epsilon_{\rm{}}}{A_l} = {\epsilon_{\rm{s}}}\left[ {\sum\limits_{n = 1}^\infty  {{L_{ln}}{B_n}}  + \sum\limits_{n = 1}^\infty  {{N_{ln}}{C_n}} } \right],
\nonumber\\
&{\epsilon_{\rm{s}}}\left[ {l{B_l} - \left( {l + 1} \right){C_l}} \right] = {\epsilon_{\rm{m}}}\left[ {l{E_l} - \left( {l + 1} \right){D_l}} \right],
\label{eqA5}
\end{align}
where
\begin{align}
{K_{ln}} &= \frac{{n!}}{{l!\left( {n - l} \right)!}}\frac{{{a^l}{d^{n - l}}}}{{{b^n}}}\left\{ {\begin{array}{*{20}{c}}
{1,\,\,\,n \geqslant l},\\
{0,\,\,\,n < l},
\end{array}} \right.
\nonumber\\
{M_{ln}} &= {\left( { - 1} \right)^{l - n}}\frac{{l!}}{{n!\left( {l - n} \right)!}}\frac{{{b^{n + 1}}{d^{l - n}}}}{{{a^{l + 1}}}}\left\{ {\begin{array}{*{20}{c}}
{1,\,\,\,n \geqslant l},\\
{0,\,\,\,n < l},
\end{array}} \right.
\nonumber\\
{L_{ln}} &= l{K_{ln}},
\nonumber\\
{N_{ln}} &=  - \left( {l + 1} \right){M_{ln}}.
\label{eqA6}
\end{align}
The relations (\ref{eqA3}), (\ref{eqA4}) follow from more general relations for harmonic functions presented in \cite{B55,B56}.

By eliminating the coefficients ${A_l}$ and ${C_l}$ from system (\ref{eqA5}), we obtain:
\begin{equation}
\label{eqA7}
\left\{ \begin{array}{l}
\left( {{\epsilon_{\rm{}}} - {\epsilon_{\rm{s}}}} \right)l\sum\limits_{n = 1}^N {{K_{ln}}{B_n}}  + \left[ {l{_{\rm{}}} + \left( {l + 1} \right){\epsilon_{\rm{s}}}} \right]\sum\limits_{n = 1}^N {{M_{ln}}{C_n}}  = 0,\\
\left( {l{\epsilon_{\rm{s}}} + \left( {l + 1} \right){\epsilon_{\rm{m}}}} \right){B_l} - \left( {l + 1} \right)\left( {{\epsilon_{\rm{s}}} - {\epsilon_{\rm{m}}}} \right){C_l} = \left( {2l + 1} \right){\epsilon_{\rm{m}}}{E_l}.
\end{array} \right.
\end{equation}
From system (\ref{eqA7}) in the dipole approximation, we obtain:
\begin{itemize}
\item[--] at $l = 1$; $n = 1,\,\,2$:
\begin{equation}
\label{eqA8}
\left\{ \begin{array}{l}
\frac{{{a^3}}}{{{b^3}}}\left( {{\epsilon_{\rm{s}}} - {\epsilon_{\rm{c}}}} \right){B_1} + \frac{{2{a^3}d}}{{{b^4}}}\left( {{\epsilon_{\rm{s}}} - {\epsilon_{\rm{c}}}} \right){B_2} - \left( {2{\epsilon_{\rm{s}}} + {\epsilon_{\rm{c}}}} \right){C_1} = 0,\\
 - \left( {{\epsilon_{\rm{s}}} + 2{\epsilon_{\rm{m}}}} \right){B_1} + 2\left( {{\epsilon_{\rm{s}}} - {\epsilon_{\rm{m}}}} \right){C_1} = 3{E_0}b{_{\rm{m}}},
\end{array} \right.
\end{equation}
\item[--] at $l = 2$; $n = 1,\,\,2$:
\begin{equation}
\label{eqA9}
\left\{ \begin{array}{l}
\left( {2{\epsilon_{\rm{s}}} + 3{\epsilon_{\rm{m}}}} \right){B_2} - 3\left( {{\epsilon_{\rm{s}}} - {\epsilon_{\rm{m}}}} \right){C_2} = 0,\\
\frac{{{a^5}}}{{{b^5}}}\left( {{\epsilon_{\rm{s}}} - {\epsilon_{\rm{c}}}} \right){B_2} - \frac{1}{2}\left( {3{\epsilon_{\rm{s}}} + 2{\epsilon_{\rm{c}}}} \right){C_2} =  - \frac{d}{b}\left( {3{\epsilon_{\rm{s}}} + 2{\epsilon_{\rm{c}}}} \right){C_1},
\end{array} \right.
\end{equation}
\end{itemize}
where
\begin{align}
\label{eqA10}
&{K_{11}} = \frac{a}{b},\quad{K_{12}} = \frac{{2ad}}{{{b^2}}},\quad{K_{22}} = \frac{{{a^2}}}{{{b^2}}},\quad{M_{11}} = \frac{{{b^2}}}{{{a^2}}},\quad{M_{21}} =  - \frac{{2{b^2}d}}{{{a^3}}},\nonumber \\
&{M_{22}} = \frac{{{b^3}}}{{{a^3}}},\quad{M_{12}} = {K_{12}} = 0.
\end{align}
Since, by definition, the dipole polarizability of a two-layer nanoparticle is proportional to the coefficient~${D_1}$,
$$
{\alpha _@} = \frac{{{D_1}}}{{b{\mathscr{E}_0}}},
$$
it follows from equations (\ref{eqA8})--(\ref{eqA10}) that we obtain formula (\ref{eq5}).

\section{Calculation of the effective relaxation rate}
\label{appendix_B}

\renewcommand*\thefigure{B.\arabic{figure}}
\setcounter{figure}{0}

\begin{figure}[htb]
\centerline{\includegraphics[width=0.65\textwidth]{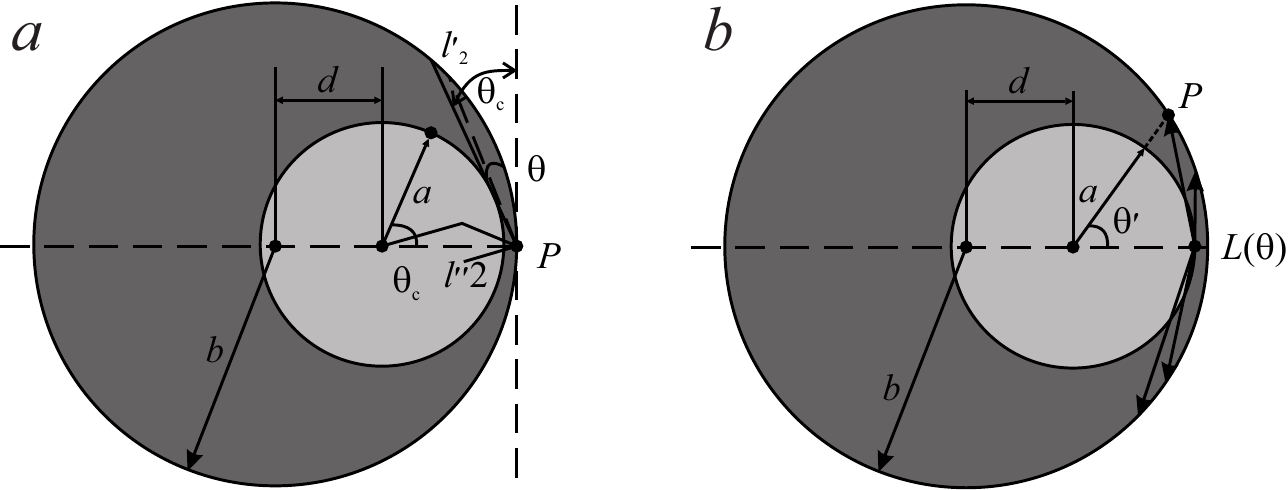}}
\caption{Schematic representation of electron trajectories when moving from the nanostructure surface to the core surface (a) and in the reverse direction (b).} \label{fig8}
\end{figure}

\textbf{Bulk relaxation.} The expression for the bulk relaxation rate is obtained by utilizing the results of the study \cite{B31} taking into account the non-concentricity and the fact that the core is a dielectric. As a result, we obtain:
\begin{equation}
\label{eqB1}
\gamma _{{\rm{bulk}}} = \frac{1}{{\left( 2 - \sin {\theta _{\rm{c}}} \right){\tau _{\rm{bulk}}}}},
\end{equation}
where $\tau _{\rm{bulk}}$ is the bulk relaxation time; $\theta _{\rm{c}}$  is the critical scattering angle (figure~\ref{fig8}a), which is determined from geometric considerations as follows:
\begin{equation}
\label{eqB2}
\cos {\theta _{\rm{c}}} = \frac{a}{{b - d}} = \frac{{\beta _{\rm{c}}^{{1 \mathord{\left/
 {\vphantom {1 3}} \right.
 \kern-\nulldelimiterspace} 3}}}}{{1 - \delta _{\rm{c}}^{{1 \mathord{\left/
 {\vphantom {1 3}} \right.
 \kern-\nulldelimiterspace} 3}}}}.
\end{equation}
Thus, the expression for the bulk relaxation rate takes the form of equation (\ref{eq11}).

\textbf{Surface relaxation.} The surface relaxation rate is determined using the approach proposed in \cite{B52}
\begin{equation}
\label{eqB3}
{\gamma _{\rm{s}}} = \frac{{{v_{\rm{F}}}}}{{{L_{{\rm{eff}}}}}},
\end{equation}
where $v_{\rm{F}}$  is the Fermi velocity of electrons; $L_{{\rm{eff}}}$  is the ``effective chord'', the average path of electrons before their first scattering event at the surface.

The effective chord is a weighted average value:
\begin{equation}
\label{eqB4}
{L_{{\rm{eff}}}} = \frac{{S{\ell _1} + {S_{\rm{c}}}{\ell _2}}}{{S + {S_{\rm{c}}}}},
\end{equation}
where ${S_{\rm{c}}} = 4\piup {a^2}$, $S = 4\piup {b^2}$  are the surface areas of the inner and outer spheres, respectively.

From expression (\ref{eqB4}), we can derive the final result:
\begin{equation}
\label{eqB5}
{L_{{\rm{eff}}}} = \frac{{{\ell _1} + \beta _{\rm{c}}^{{2 \mathord{\left/
 {\vphantom {2 3}} \right.
 \kern-\nulldelimiterspace} 3}}{\ell _2}}}{{1 + \beta _{\rm{c}}^{{2 \mathord{\left/
 {\vphantom {2 3}} \right.
 \kern-\nulldelimiterspace} 3}}}}.
\end{equation}
In formulas (\ref{eqB4}) and (\ref{eqB5}), the quantities ${\ell _1}$ and ${\ell _2}$ represent the electron trajectory lengths as they move toward the nanostructure and core surfaces (figure~\ref{fig8}a).

Now, we compute the trajectory lengths ${\ell _1}$ and ${\ell _2}$. According to figure~\ref{fig8}a, we obtain:
\begin{equation}
\label{eqB6}
{\ell _1} = {\ell '_1} + {\ell ''_1},
\end{equation}
where
\begin{equation}
\label{eqB7}
{\ell '_1} = \int\limits_0^{{\theta _{\rm{c}}}} {{{L'}_1}\left( \theta  \right)\cos \theta \,\rd\theta },
\end{equation}
\begin{equation}
\label{eqB8}
{\ell ''_1} = \int\limits_{{\theta _{\rm{c}}}}^{{\piup }/{2}} {{{L''}_1}\left( \theta  \right)\cos \theta \,\rd\theta }.
\end{equation}
From geometric considerations, we obtain:
\begin{equation}
\label{eqB9}
{L'_1}\left( \theta  \right) = 2b\sin \theta,
\end{equation}
\begin{equation}
\label{eqB10}
{L''_1}\left( \theta  \right) = \left( {b - d} \right)\sin \theta  - \sqrt {{a^2} - {{\left( {b - d} \right)}^2}{{\cos }^2}\theta }.
\end{equation}
Substituting (\ref{eqB9}) and (\ref{eqB10}) into (\ref{eqB7}) and (\ref{eqB8}) and evaluating the corresponding integrals, we obtain:
\begin{align}
\label{eqB11}
{\ell _1} = \,&b\left\{ {1 - \frac{{\beta _{\rm{c}}^{{2}/{3}}}}{{{{\left( {1 - \delta _{\rm{c}}^{{1}/{3}}} \right)}^2}}} + \frac{1}{2}\left( {1 - \delta _{\rm{c}}^{{1}/{3}}} \right)\left[ {1 - \left( {1 - \frac{{\beta _{\rm{c}}^{{2}/{3}}}}{{{{\left( {1 - \delta _{\rm{c}}^{{1}/{3}}} \right)}^2}}}} \right)} \right.} \right. 
\nonumber\\
&\times \left. {\left. {\left( {\frac{1}{2}\ln \frac{{1 - \beta _{\rm{c}}^{{1}/{3}} - \delta _{\rm{c}}^{{1}/{3}}}}{{1 + \beta _{\rm{c}}^{{1}/{3}} - \delta _{\rm{c}}^{{1}/{3}}}} + \frac{{\beta _{\rm{c}}^{{1}/{3}}\left( {1 - \delta _{\rm{c}}^{{1}/{3}}} \right)}}{{{{\left( {1 - \delta _{\rm{c}}^{{1}/{3}}} \right)}^2} - \beta _{\rm{c}}^{{2}/{3}}}}} \right)} \right]} \right\}.
\end{align}
According to figure~\ref{fig8}b, we obtain
\begin{equation}
\label{eqB12}
{\ell _2} = \int\limits_0^{{\theta _{\rm{c}}}} {{L_2}\left( {\theta '} \right)\sin \theta' \,\rd\theta '}.
\end{equation}
Introducing a new variable $x = \cos \theta '$ we obtain:
\begin{equation}
\label{eqB13}
{\ell _2} = \int\limits_{\cos {\theta _{\rm{c}}}}^1 {\frac{{\left( {b - d} \right)\rd x}}{{x - \cos {\theta _{\rm{c}}}}}\left[ {x - \frac{1}{{2\left( {\lambda  - x} \right)}} + \frac{{{x^2}}}{{2\left( {\lambda  - x} \right)}}} \right]},
\end{equation}
where
\begin{equation}
\label{eqB14}
\lambda  = \frac{{\beta _{\rm{c}}^{{2 \mathord{\left/
 {\vphantom {2 3}} \right.
 \kern-\nulldelimiterspace} 3}} + {{\left( {1 - \delta _{\rm{c}}^{{1 \mathord{\left/
 {\vphantom {1 3}} \right.
 \kern-\nulldelimiterspace} 3}}} \right)}^2}}}{{2\beta _{\rm{c}}^{{1 \mathord{\left/
 {\vphantom {1 3}} \right.
 \kern-\nulldelimiterspace} 3}}\left( {1 - \delta _{\rm{c}}^{{1 \mathord{\left/
 {\vphantom {1 3}} \right.
 \kern-\nulldelimiterspace} 3}}} \right)}}.
\end{equation}
Evaluating the integral in equation (\ref{eqB13}), we obtain:
\begin{equation}
\label{eqB15}
{\ell _2} = \frac{b}{2}\left( {1 - \delta _{\rm{c}}^{{1 \mathord{\left/
 {\vphantom {1 3}} \right.
 \kern-\nulldelimiterspace} 3}}} \right)\left[ {1 - \frac{{\beta _{\rm{c}}^{{1 \mathord{\left/
 {\vphantom {1 3}} \right.
 \kern-\nulldelimiterspace} 3}}}}{{1 - \delta _{\rm{c}}^{{1 \mathord{\left/
 {\vphantom {1 3}} \right.
 \kern-\nulldelimiterspace} 3}}}} - \frac{{{{\left( {1 - \delta _{\rm{c}}^{{1 \mathord{\left/
 {\vphantom {1 3}} \right.
 \kern-\nulldelimiterspace} 3}}} \right)}^2} - \beta _{\rm{c}}^{{2 \mathord{\left/
 {\vphantom {2 3}} \right.
 \kern-\nulldelimiterspace} 3}}}}{{2\beta _{\rm{c}}^{{1 \mathord{\left/
 {\vphantom {1 3}} \right.
 \kern-\nulldelimiterspace} 3}}\left( {1 - \delta _{\rm{c}}^{{1 \mathord{\left/
 {\vphantom {1 3}} \right.
 \kern-\nulldelimiterspace} 3}}} \right)}}\ln \frac{{1 - \beta _{\rm{c}}^{{1 \mathord{\left/
 {\vphantom {1 3}} \right.
 \kern-\nulldelimiterspace} 3}} - \delta _{\rm{c}}^{{1 \mathord{\left/
 {\vphantom {1 3}} \right.
 \kern-\nulldelimiterspace} 3}}}}{{1 + \beta _{\rm{c}}^{{1 \mathord{\left/
 {\vphantom {1 3}} \right.
 \kern-\nulldelimiterspace} 3}} - \delta _{\rm{c}}^{{1 \mathord{\left/
 {\vphantom {1 3}} \right.
 \kern-\nulldelimiterspace} 3}}}}} \right].
\end{equation}
Substituting (\ref{eqB11}) and (\ref{eqB15}) into (\ref{eqB5}), we obtain
\begin{align}
\label{eqB16}
{L_{{\rm{eff}}}} =&\, \frac{b}{{1 + \beta _{\rm{c}}^{{2}/{3}}}}\left\{ {1 - \frac{{\beta _{\rm{c}}^{{2}/{3}}}}{{{{\left( {1 - \delta _{\rm{c}}^{{1}/{3}}} \right)}^2}}} + \frac{1}{2}\left( {1 - \delta _{\rm{c}}^{{1}/{3}}} \right)\left( {1 + \beta _{\rm{c}}^{{2}/{3}}} \right)\left( {1 - \frac{{\beta _{\rm{c}}^{{1}/{3}}}}{{1 - \delta _{\rm{c}}^{{1}/{3}}}}} \right)} \right. 
\nonumber\\
 -& \left. {\frac{1}{4}\left( {1 - \delta _{\rm{c}}^{{1}/{3}}} \right)\left[ {1 - \frac{{\beta _{\rm{c}}^{{2}/{3}}}}{{{{\left( {1 - \delta _{\rm{c}}^{{1}/{3}}} \right)}^2}}} + \beta _{\rm{c}}^{{1}/{3}}\frac{{{{\left( {1 - \delta _{\rm{c}}^{{1}/{3}}} \right)}^2} - \beta _{\rm{c}}^{{2}/{3}}}}{{1 - \delta _{\rm{c}}^{{1}/{3}}}}} \right]\ln \frac{{1 - \beta _{\rm{c}}^{{1}/{3}} - \delta _{\rm{c}}^{{1}/{3}}}}{{1 + \beta _{\rm{c}}^{{1}/{3}} - \delta _{\rm{c}}^{{1}/{3}}}}} \right\}.
\end{align}
Thus, the final expression for the surface relaxation rate takes the form of equation (\ref{eq12}).

\textbf{Radiative damping.} Now, we determine the rate of radiative damping.

Here, it is necessary to consider two cases: $\ell _{{\rm{bulk}}}^s > 2{t_{{\rm{eff}}}}$ and $\ell _{{\rm{bulk}}}^s \leqslant 2{t_{{\rm{eff}}}}$, where $\ell _{{\rm{bulk}}}^s = {v_{\rm{F}}}{\tau _{{\rm{bulk}}}}$ is the effective thickness of the metallic shell:
\begin{equation}
\label{eqB17}
{t_{{\rm{eff}}}} = \frac{1}{\piup }\int\limits_0^\piup  {t\left( \theta  \right)\,\rd\theta }.
\end{equation}
Since
$$
t\left( \theta  \right) = b - a - d\cos \theta,
$$
by evaluating the integral in (\ref{eqB17}), we obtain
\begin{equation}
\label{eqB18}
{t_{{\rm{eff}}}} = b\left( {1 - \beta _{\rm{c}}^{{1 \mathord{\left/
 {\vphantom {1 3}} \right.
 \kern-\nulldelimiterspace} 3}} - \frac{2}{\piup }\delta _{\rm{c}}^{{1 \mathord{\left/
 {\vphantom {1 3}} \right.
 \kern-\nulldelimiterspace} 3}}} \right).
\end{equation}
For $\ell _{{\rm{bulk}}}^s > 2{t_{{\rm{eff}}}}$ we have:
\begin{equation}
\label{eqB19}
{\gamma _{{\rm{rad}}}} = \frac{2}{{9\epsilon_{\rm{m}}^{{1 \mathord{\left/
 {\vphantom {1 2}} \right.
 \kern-\nulldelimiterspace} 2}}}}{\left( {\frac{{{\omega _p}}}{c}} \right)^3}{V_{\rm{s}}}\frac{{{v_{\rm{F}}}}}{{{L_{{\rm{eff}}}}}}
\end{equation}
and for $\ell _{{\rm{bulk}}}^s \leqslant 2{t_{{\rm{eff}}}}$
$$
{\gamma _{{\rm{rad}}}} = \frac{2}{{9\epsilon_{\rm{m}}^{{1 \mathord{\left/
 {\vphantom {1 2}} \right.
 \kern-\nulldelimiterspace} 2}}}}{\left( {\frac{{{\omega _p}}}{c}} \right)^3}{V_{\rm{s}}}{\left[ {\int\limits_0^{{\theta _{\rm{c}}}} {{\tau _{{\rm{bulk}}}}\cos \theta \,\rd\theta }  + 2\int\limits_{{\theta _{\rm{c}}}}^{{\piup }/{2}} {{\tau _{{\rm{bulk}}}}\cos \theta \,\rd\theta } } \right]^{ - 1}}
$$
or
\begin{equation}
\label{eqB20}
{\gamma _{{\rm{rad}}}} = \frac{2}{{9\epsilon_{\rm{m}}^{{1 \mathord{\left/
 {\vphantom {1 2}} \right.
 \kern-\nulldelimiterspace} 2}}}}{\left( {\frac{{{\omega _p}}}{c}} \right)^3}{V_{\rm{s}}}\frac{1}{{\left( {2 - \sin {\theta _{\rm{c}}}} \right){\tau _{{\rm{bulk}}}}}}.
\end{equation}
Taking into account that
$$
{V_{\rm{s}}} = V - {V_0} = \left( {1 - {\beta _{\rm{c}}}} \right)V,
$$
where ${V_0} = {{4\piup {a^3}}}$, as well as using equations (\ref{eqB1}) and (\ref{eqB3}) instead of (\ref{eqB19}) and (\ref{eqB20}), we obtain expression~(\ref{eq13}).


%
%

\ukrainianpart

\title{Поглинання та розсіяння світла метал-діелектричними нанояйцями}
\author{Р. Ю.~Корольков\refaddr{label1}, Р. О.~Малиш\refaddr{label1}, А. В.~Коротун\refaddr{label1,label2}, Р. А.~Куликовський\refaddr{label1}}
\addresses{
\addr{label1} Національний університет ``Запорізька політехніка'', 69011 Запоріжжя, Україна
\addr{label2} Інститут металофізики ім. Г.В. Курдюмова НАН України, 03142 Київ, Україна
}
%
%
%

\makeukrtitle

\begin{abstract}
\tolerance=3000%
У роботі досліджено оптичні та плазмонні властивості метал-діелектричних нанояєць. Визначено частотні залежності поляризовності, перерізів поглинання і розсіювання та радіаційної ефективності. Отримано вирази для розмірних залежностей частот поверхневого плазмонного резонансу. Встановлено причини синіх та червоних зсувів максимумів поляризованості та перерізів поглинання і розсіювання, а також змін їх кількості та амплітуди. Надано рекомендації щодо використання у різних спектральних діапазонах матеріалів, що мають максимальну радіаційну ефективність.
\keywords поверхневий плазмонний резонанс, нанояйце, поляризовність, перерізи поглинання і розсіювання, діелектрична функція, ефективна швидкість релаксації, радіаційна ефективність

\end{abstract}

\end{document}